\documentclass[final,3p,times,twocolumn]{elsarticle}
\usepackage{graphicx}
\usepackage{amsmath}
\usepackage{amssymb}
\usepackage{xcolor}
\usepackage{hyperref}

\journal{Surfaces and Interfaces}

\begin{document}

\title{Interface magnetic coupling and magnetization dynamic of La$_{2/3}$Sr$_{1/3}$MnO$_3$ single layer and (La$_{2/3}$Sr$_{1/3}$MnO$_3$/SrRuO$_3$)$_n$ (n = 1, 5) superlattice on SrTiO$_3$(001) substrate}

\author{Ilyas Noor Bhatti}\ead{inoorbhatti@gmail.com}\address{Department of Physics, Jamia Millia Islamia University, 110025 New Delhi, India}
\author{Rachna Chaurasia}\address{Department of Physics, Ramjas College, Delhi University, 110007 New Delhi, India}
\author{Kazi Rumanna Rahman}\address{Fuel Cell Institute, University Kebangsaan Malaysia, Bangi 43600, Selangor, Malaysia}
\author{Sukhendu Sadhukhan}\address{SATIE, ENS Paris-Saclay, CNRS, Universit$\acute{e}$ Paris-Saclay, 91190 Gif-Sur-Yvette, France}
\author{Amantulla Mansuri}\address{Department of Physics, Govt. Autonomous PG College, Chhindwara, 480001, M.P. India}
\author{Imtiaz Noor Bhatti}\ead{inbhatti07@gmail.com}\address{Laboratoire Albert Fert, CNRS, Thales, Universit$\acute{e}$ Paris-Saclay, 1 avenue Augustin Fresnel, 91767 Palaiseau, France}

\begin{abstract}
In this work, we investigate the structural, magnetic, and microwave magnetic dynamics of multilayered \([{\rm LSMO}/{\rm SRO}]_n\) heterostructures \((n = 1 \text{ and } 5)\) grown on SrTiO\(_3\) (001) substrates. X-ray diffraction confirms high crystallinity and atomically sharp interfaces. Magnetic measurements reveal strong interfacial magnetic coupling, with a distinct two-step magnetization switching observed in the \(n = 5\) heterostructure, while this feature is significantly suppressed in the \(n = 1\) structure. Ferromagnetic resonance (FMR) analysis shows a broad linewidth, pronounced positive magnetic anisotropy, and Gilbert damping on the order of \(10^{-2}\), with damping decreasing as the number of multilayer repetitions increases. These observations demonstrate that Ru--Mn exchange coupling at the interface critically governs the magnetic response and dynamic behavior of the system. The tunable switching and damping properties highlight such oxide heterointerfaces as promising platforms for exploring spin textures, magnetic domain behavior, and room-temperature spintronic applications.
\end{abstract}


\maketitle

\section{Introduction}
In recent years, artificially constructed interfaces between dissimilar materials have attracted significant attention from researchers due to their ability to host exotic states different from their bulk counterparts. Such engineered interfaces have revealed novel properties in various oxide heterostructures.\cite{moko, huang, lei} The strong electron correlations at oxide interfaces, coupled with the interplay between charge, spin, orbital, and structural degrees of freedom, provide opportunities to realize tunable properties.\cite{zhang, van, ja, apchen, yli, lzhe} Understanding and effectively controlling these coupling effects are essential steps toward engineering emergent interfacial phases with tailored functionalities. Thus, interfaces offer an additional degree of freedom, enabling the emergence of diverse physical phenomena and inspiring innovative concepts for advanced electronic and magnetic devices. Recently, dimensional tuning of interface coupling–induced ferromagnetism has been explored in CaRuO$_3$/SrCuO$_2$ superlattices, where charge transfer at the interface was found to differ between chain- and plane-type SrCuO$_2$ interfaces, enabling precise tuning of the interfacial magnetic state through dimensional control.\cite{lzhe} An interfacial Dzyaloshinskii–Moriya interaction (DMI)–driven emergent magnetoelectric phase transition has been realized in Sr$_2$IrO$_4$/SrTiO$_3$ superlattices engineered with artificial ferroelectricity~\cite{xliu}. In another study, involving paramagnetic LaNiO$_3$ and ferromagnetic La$_{2/3}$Ca$_{1/3}$MnO$_3$ films deposited on an STO (001) substrate, results revealed that interfacial interactions can induce complex charge and spin structures in the otherwise paramagnetic LaNiO$_3$.\cite{soltan} 

The coupling between magnetic layers across interfaces is a key property for magnetic sensors, spintronic devices, and data storage technologies. Among various oxides, ferromagnetic half-metallic manganites are particularly promising for spintronic and multifunctional applications. Heterostructures such as La$_{2/3}$Sr$_{1/3}$MnO$_3$/BaFiO$_3$, La$_{2/3}$Sr$_{1/3}$MnO$_3$/SrIrO$_3$, and La$_{2/3}$Sr$_{1/3}$MnO$_3$/SrRuO$_3$ have been extensively investigated over the past decade.\cite{halder, nara, helen, ravi, luis} Among these, La$_{2/3}$Sr$_{1/3}$MnO$_3$ and SrRuO$_3$ stand out due to their excellent epitaxial compatibility and lattice matching, enabling high-quality heterostructures that display a range of interface-driven magnetic phenomena. A one-band double-exchange model studied through Monte Carlo simulations for the La$_{2/3}$Sr$_{1/3}$MnO$_3$/SrRuO$_3$ superlattice has revealed antiferromagnetic interfacial coupling and a three-step magnetization switching in the hysteresis loop \cite{halder}. Furthermore, La$_{2/3}$Sr$_{1/3}$MnO$_3$/SrRuO$_3$ multilayers grown on SrTiO$_3$ (001) substrates exhibit a positive exchange bias and asymmetric magnetic reversal, attributed to two distinct interfacial antiferromagnetic coupling strengths, as described by an extended Stoner–Wohlfarth model \cite{solignac,ziese1,ziese2}. Despite extensive studies on La$_{2/3}$Sr$_{1/3}$MnO$_3$/SrRuO$_3$ multilayers, the behavior of superlattices with an increased number of interfaces remains largely unexplored.
\begin{figure*}[tb]
	\centering
		\includegraphics[width=16cm]{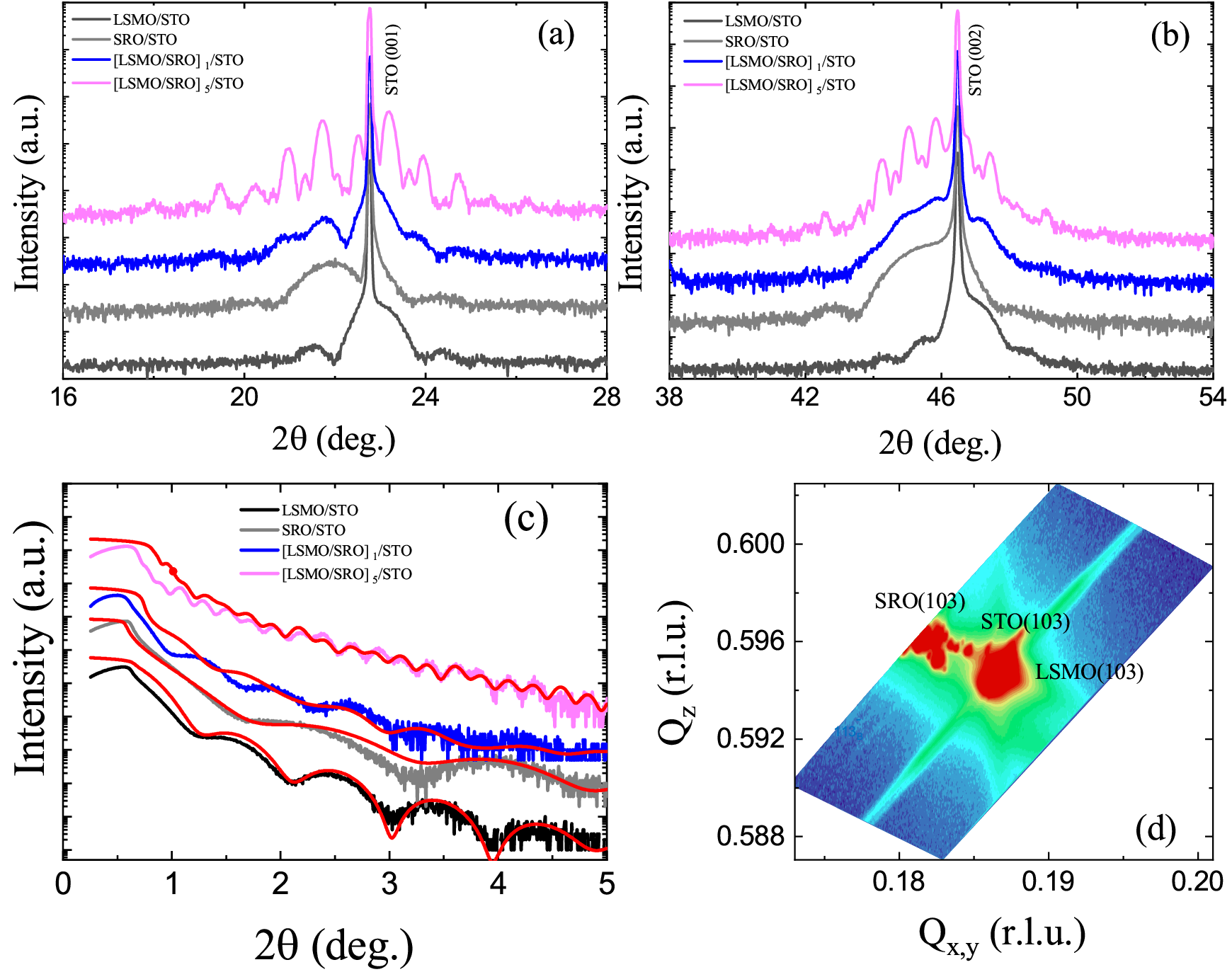}
	\caption{(color online) X-ray characterization (a-b) shows the (001) and (002) peaks of 2$\theta$/$\omega$ scans respectively, here for superlattice sample multipal peaks associated with interfaces are also seen where numbering represent layers (c) X-ray reflectivity along with simulated curves (solid red lines) are presented. (d) Reciprocal space mapping near (103) reflection}
	\label{fig:Fig1}
\end{figure*}
In the present study, we investigate 3$d$/4$d$-based [La$_{2/3}$Sr$_{1/3}$MnO$_3$/SrRuO$_3$]$_n$ multilayer heterostructures (where $n = 1$ and $5$) epitaxially grown on SrTiO$_3$ (STO) substrates. SrRuO$_3$ (SRO), a 4$d$-based itinerant ferromagnet, belongs to the ruthenate series Sr$_{3n+1}$Ru$_n$O$_{3n+1}$ (with $n = \infty$ for SRO). In its bulk form, SRO crystallizes in an orthorhombic perovskite structure and exhibits paramagnetic behavior at room temperature, becoming ferromagnetic below its Curie temperature ($T_c \approx 160$~K).\cite{renu1, renu2} The magnetic and structural characteristics of SRO make it an excellent conducting oxide and a widely used electrode material in oxide-based devices. On the other hand, La$_{2/3}$Sr$_{1/3}$MnO$_3$ (LSMO) is a 3$d$-based half-metallic manganite, well-known for its high spin polarization and strong double-exchange-driven ferromagnetism. It crystallizes in a distorted perovskite structure and possesses a relatively high Curie temperature ($T_c \approx 370$~K).\cite{snyder} When these two materials LSMO and SRO are combined in a multilayer configuration, they form a highly coherent and lattice-matched interface due to their structural compatibility. However, at the interface, complex interactions arise due to the competition between the 3$d$–4$d$ electronic orbitals, exchange interactions, and magnetocrystalline anisotropy, leading to the emergence of novel magnetic behaviors not present in their bulk counterparts. In this work, we focus on understanding how the magnetic coupling across the LSMO/SRO interface evolves as the number of interfaces ($n$) increases. Such an investigation provides valuable insight into interface-driven magnetism in complex oxide systems and contributes to the broader understanding required for designing advanced spintronic and magnetoresistive devices based on transition-metal oxide heterostructures.

\begin{figure}[h!]
	\centering
		\includegraphics[width=8cm]{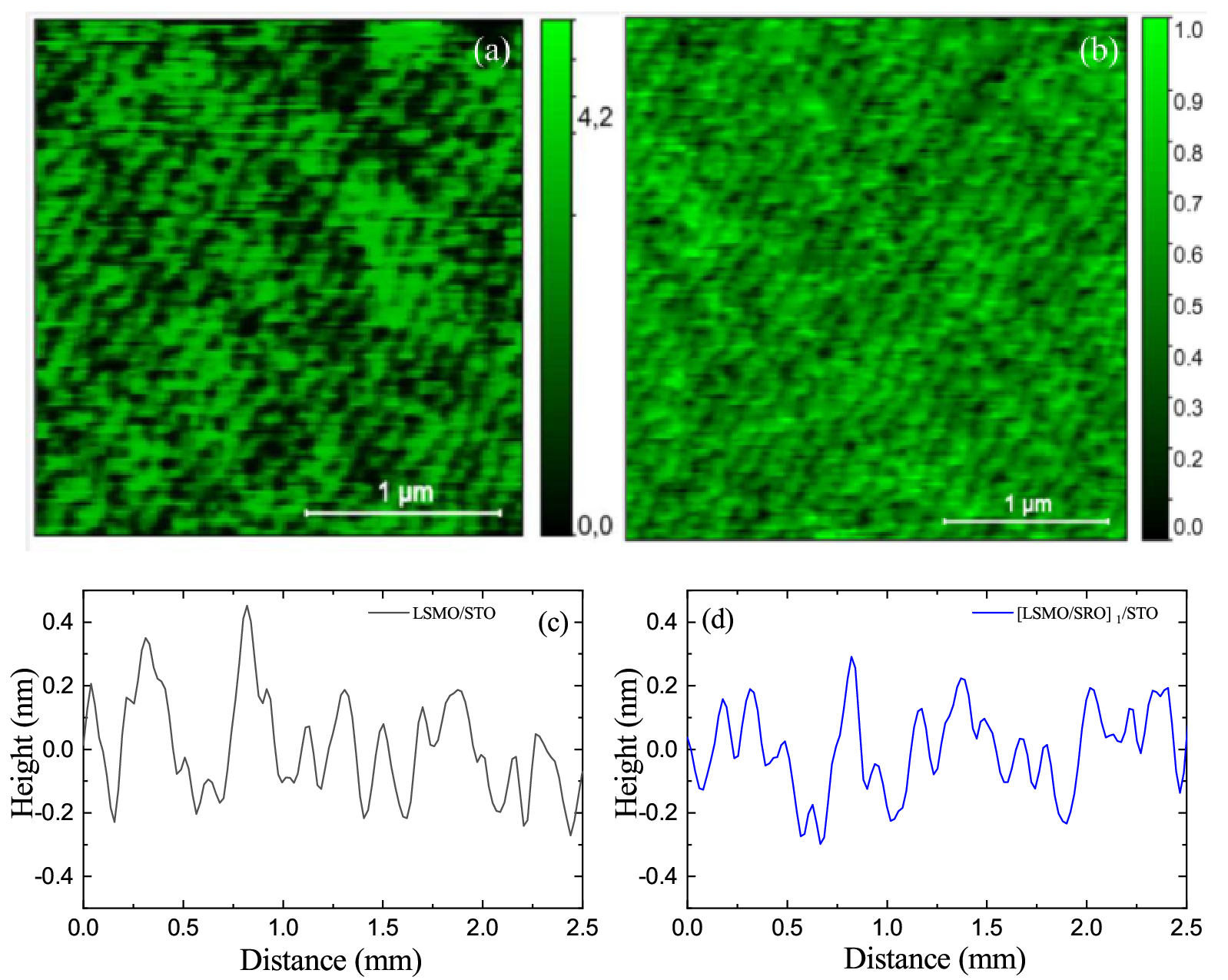}
	\caption{(color online) Atomic force microscope surface micrograph (a) shows AFM image for LSMO/STO and (b) shows image for [LSMO/SRO]$_1$/STO bi-layer sample. (c and d) present the line profile along a arbitrary line for each image.}
	\label{fig:Fig2}
\end{figure}

\section{Experimental Details}
The epitaxial thin films, heterostructures, and superlattice of La$_{2/3}$Sr$_{1/3}$MnO$_3$ (LSMO) and SrRuO$_3$ (SRO) were deposited on SrTiO$_3$ (STO) (001) substrates. The high-quality and chemically pure PLD targets of LSMO and SRO were commercially procured from Toshiba Materials Co., Ltd. We used 5 $\times$ 5 mm$^2$ STO substrates for sample growth. Substrates were first cleaned with acetone and then with isopropanol in ultrasonic bath for 5 minutes each. For further cleaning, the substrate was heated to 900 $^{\circ}$C at 0.4 mbar in oxygen pressure to remove remaining organic impurities. Thin film deposition was carried out using an Nd-YAG laser based Pulsed laser deposition system, with wavelength $\lambda$ = 295 nm, frequency 2.5 Hz, 0.26 mbar oxygen pressure, and a laser fluence of approximately $\sim$2.2 mJ/cm$^2$. The substrates was kept at a constant temperature of 650 $^o$C during deposition, and after growth the samples were cooled to ambient temperature under control cooling rate of 20 $^o$C/min in oxygen pressure of 300 mbar. The deposited samples were characterized for structural, surface morphology, and magnetic properties.  Structural characterization, x-ray diffraction, x-ray reflectivity, and reciprocal space mapping were performed using a PANalytical differectometer, equipped with monochromatic Cu k$\alpha_1$ ($\lambda$ = 1.540 $\AA$) radiation. Surface morphology and roughness were estimated from AFM images obtained using an atomic force microscope (Innova, Bruker). The elemental composition and chemical states of the films were analyzed using an X-ray photoelectron spectroscopy. The measurements were carried out in-situ immediately after film deposition using the XPS unit integrated with the growth chamber to prevent surface exposure and contamination. X-ray photoelectron spectroscopy (XPS) was performed using a x-ray photoelectron emission spectrometer from Omicron Nanotechnology equipped with a monochromatic a Mg K$\alpha$ X-ray source (h$\mu$ = 1253.6 eV) operated at 30 kV and 20 mA emission current. Spectra were collected at a pass energy of 20 eV in normal emission geometry. There was charging effect during measurement to address this effect in final data the binding energy scale was calibrated by referencing the adventitious carbon (C 1s) peak to 284.8 eV. The data is collected with a step size of 50 meV for all core level spectrum. Spectral analysis was performed using CasaXPS software. A Shirley-type background was applied to all core-level spectra to account for inelastic electron scattering. The peaks were fitted with a product of Gaussian and Lorentzian functions for all components. The number of components for each core level was determined based on expected chemical states, and the need to achieve a satisfactory fit with minimal residuals.
To examine the effect of interfacial structure on magnetic coupling, field-dependent magnetization measurements were performed at various temperatures. DC magnetization was measured using a SQUID magnetometer (Quantum Design, USA), with all samples cooled under zero magnetic field (ZFC) prior to measurement. Room temperature magnetization dynamics in the microwave frequency regime were investigated using a home-built coplanar waveguide-based ferromagnetic resonance (FMR) setup. During FMR measurements, the samples were positioned on the waveguide with their plane aligned parallel to the applied external magnetic field. A small excitation AC magnetic field was applied, and by sweeping the external DC magnetic field at a fixed microwave frequency, the FMR absorption spectra were recorded when the resonance condition was satisfied.

\begin{figure*}[tb]
	\centering
		\includegraphics[width=16cm]{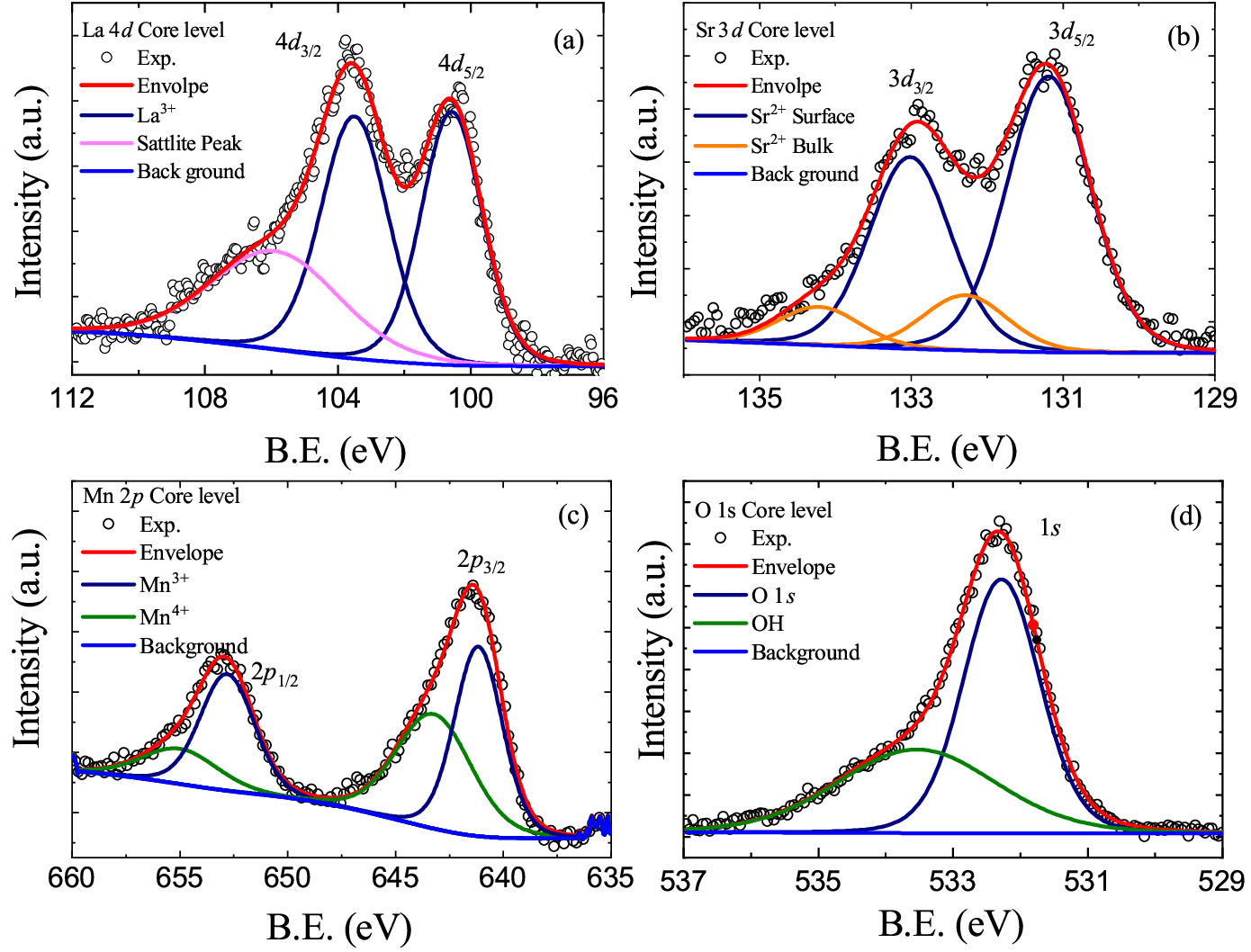}
\caption{(Color online) X-ray Photoemission spectroscopy (a) shows the core level spectra of La 4$d$ (b) shows the core level spectra of Sr 3$d$ (c)  shows the core level spectra of Mn 2$p$ (d) shows the core level spectra of  O 1$s$.  In the figure the red solid line is the overall envelop of the XPS spectrum and the other colored solid lines are the respective fitted peaks as marked in figures.}
	\label{fig:Fig3}
\end{figure*}

\begin{figure}[h!]
	\centering
		\includegraphics[width=8cm]{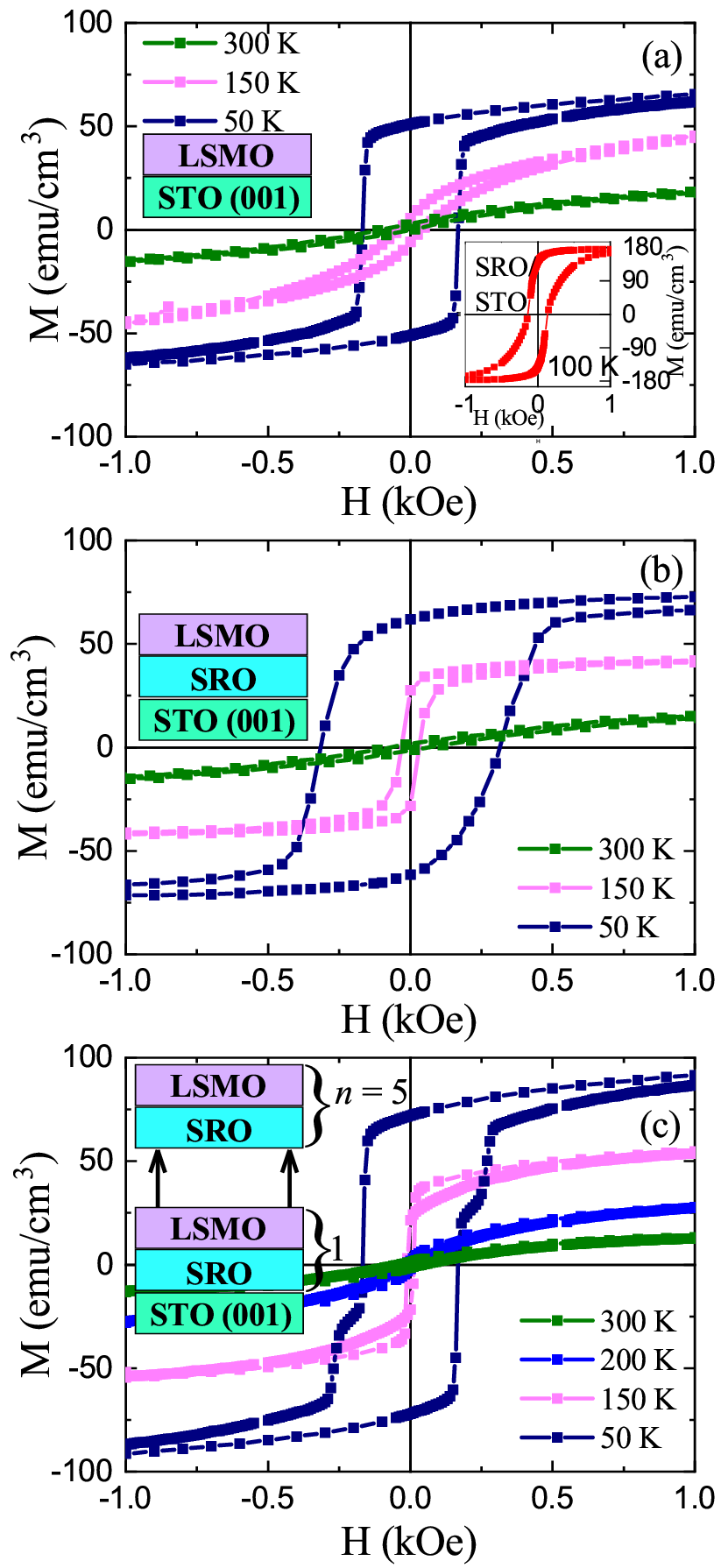}
	\caption{(color online) Isothermal magnetization measured at 300 K, 150 K and 50 K presented for (a) LSMO/STO, inset shows Isothermal magnetization of SRO at 100 K, (b) [LSMO/SRO]$_1$, and (c) [LSMO/SRO]$_5$. Inset of each figure present the layer schematic of grown samples.}
	\label{fig:Fig4}
\end{figure}

\section{Results and Discussion}

\subsection{Structural and Surface Characterization}
Fig. 1 illustrates the structural characterization of SrRuO$_3$/STO, LSMO/STO, and [LSMO/SRO]$_n$/STO ($n=1$ \& $5$) Films. Bulk SrRuO$_3$ is realized from the Ruddlesden-Popper series Sr$_{n+1}$Ru$_n$O$_{3n+1}$, with $n = \infty$, representing an infinite stacking of SrRuO$_3$ layers, forming a three-dimensional structural network. Bulk SrRuO$_3$ adopts the orthorhombic crystal structure with \textit{Pbnm} space group. The lattice parameters of the orthorhombic unit cell are $a = 5.53$~\AA, $b = 5.57$~\AA, and $c = 7.85$~\AA. In the unstrained orthorhombic phase, the Ru-O bonds are approximately two times shorter than the Sr-O bonds, inducing a distortion in the RuO$_6$ octahedra. As the temperature increases, the orthorhombic structure transitions to tetragonal and subsequently to cubic structures. The pseudocubic lattice parameter, calculated from the bulk lattice parameters, is approximately:
\[
a_\text{pc} \approx \sqrt{\frac{a^2 + b^2}{2}}, \quad a_\text{pc} = 3.93~\text{\AA}.
\]
For the STO substrate, the pseudocubic lattice parameter is $3.90$~\AA, which results in a lattice mismatch of $0.7 \%$ with SrRuO$_3$. Similarly, the pseudocubic lattice parameter of LSMO is $3.88$~\AA, leading to a lattice mismatch of $0.5 \%$ with STO. Figs. 1(a) and 1(b) show the $\theta$-$2\theta$ scan of x-ray diffraction (XRD) plots for the (001) and (002) reflections of SrRuO$_3$, LSMO, and [LSMO/SRO]$_n$ deposited on STO. The XRD patterns indicate a crystalline phase without any trace of impurity or additional phases. A satellite peak near the substrate peak exhibits the epitaxial growth of the film, as shown in Figs. 1(a) and 1(b). Due to the close match in pseudocubic lattice parameters between SrRuO$_3$ and STO substrate, LSMO and the STO substrate, the corresponding satellite peaks overlap with the STO Bragg reflection. As a result, resolving the individual out-of-plane lattice parameters of the films becomes challenging, and the strain state is difficult to determine from the XRD data. The XRD scans of the multilayer [LSMO/SRO]$_n$ ($n=1$ \& $5$) are shown in Figs. 1a and 1b. For $n = 1$, distinct peaks corresponding to LSMO and SRO are observed on either side of the substrate peak. The superlattice satellite peaks appear symmetrically on both sides of the substrate reflection. These peaks are more pronounced in the [LSMO/SRO]$_5$ structure due to the increased periodicity of the multilayer. Their presence confirms the well-defined periodic modulation of the lattice. Fig. 1(c) presents the x-ray reflectivity (XRR) curve of SrRuO$_3$, LSMO, and [LSMO/SRO]$_n$ ($n=1$ \& $5$). The presence of Kiessig fringes across the entire thickness range indicates well-defined interfaces and layered structures in the films. From the x-ray reflectivity data, we have calculated the thickness. The thickness for LSMO/STO, SRO/STO, [SRO/LSMO]$_1$, and [LSMO/SRO]$_5$ are 11~nm, 9~nm, 21.5~nm, and 97~nm, respectively. Fig. 1(d) shows the x-ray diffraction reciprocal space mapping (RSM) near the (103) reflection, confirming the coherent growth of the films on the STO substrate.

Fig 2 shows the AFM study for representative samples. Figs. 2a and 2b show micrograph of the LSMO/STO and [LSMO/SRO]$_1$ respectively. The 2.5$\times$2.5 $\mu$$m^2$ image was obtained at room temperature in tapping mode for all samples. We have observed that the samples have smooth surface and roughness (R$_{rms}$) is in the ranges of $\pm$0.3-0.4 nm. [LSMO/SRO]$_5$ may have higher roughness sue to increase in periodicity and thickness. The line profile have been extracted along a random line through the center of the image and presented in the Figs. 2c and 2d, which clearly evidence the samples are deposited with good quality and smooth surfaces.

\subsection{X-ray photo-electron spectroscopy (XPS)}
The oxidation states of the cations in a material play significant role in determining its physical properties. In this study, we utilized x-ray photoelectron spectroscopy to determine the charge state of La, Sr, Mn cations in La$_{2/3}$Sr$_{1/3}$MnO$_3$ grown on STO (001) substrate. Fig. 3 shows the spectra of  La, Sr, Mn cations. In Fig. 3(a) La 4$d$ core level spectrum is shown. The La 4$d$ spectrum was fitted with a main spin-orbit doublet (4$d$$_{5/2}$ and 4$d$$_{3/2}$) and a single satellite peak at higher binding energy which is  in good agreement with the literature.\cite{Ji} Two peaks are show one at 100 eV and second peak at 103.8 eV corresponding to La 4$d_{3/2}$ and La 4$d_{5/2}$  respectively. The resultant spin-orbit splitting energy of La 4$d$ is 3.8 eV. This satellite peak is a characteristic feature of La$^{3+}$ in oxides, commonly associated with a screened final state or configuration interaction, and is included to accurately represent the intrinsic electronic structure of the material. The core level peak position of La 4$d$ in this spectra are evidently corresponds to +3 oxidation state of La cations.
 
The Sr 3$d$  core level spectrum  is shown in Fig. 3(b). Two distinct peaks are present in the XPS spectrum, one at 131.2 eV and the other at 133 eV, resulting from the splitting of Sr 3$d$ into 3$d$$_{3/2}$ and 3$d$$_{5/2}$ with a splitting energy of 1.8 eV in good agreement with the literature\cite{imtiaz1,imtiaz2,imtiaz3} However, careful deconvolution of Sr 3$d$ spectrum reveals surface and bulk contribution of Sr cations. The Sr 3$d$ core level was fitted with two spin-orbit doublets to account for the distinct electronic environments of bulk and surface Sr species. The bulk doublet corresponds to Sr$^{2+}$ ions within the perovskite lattice, while the surface doublet, shifted to higher binding energy by approximately 1.0 eV, is attributed to surface termination and/or the formation of Sr carbonate or hydroxide species due to air exposure. For both doublets, the spin-orbit splitting was fixed at 1.8 eV, the branching ratio was constrained to 3:2, and the FWHM was kept equal for all components.\cite{ji1,dana,nardi}  These peaks (3$d$$_{3/2}$ and 3$d$$_{5/2}$) confirm the +2 oxidation state of Sr cations.
The core level spectrum of Mn 2$p$ is presented in Fig. 3(c). Deconvolution of core level Mn 2$p$ spectrum reveals two set of Mn 2$p$ XPS peaks splitted into Mn 2$p$$_{3/2}$ and 2$p$$_{1/2}$ at 641.1 eV and 652.7 eV corresponding to Mn$^{3+}$ and peaks at 643.3 eV and 654.8 eV  corresponding to Mn$^{4+}$. \cite{ilyas1, ilyas2} This deconvolution of core level Mn 2$p$ spectrum confirms a mixed valence state of Mn$^{3+}$ and Mn$^{4+}$. While fitting the Mn 2$p$ core level, we have set full width half maximum (FWHM) as free variable since multiplet splitting and shake-up satellites differ with oxidation state and core hole lifetime broadening depends on the valence electron configuration also different local coordination or hybridization can broaden or sharpen peaks differently, so different line width is possible and gives a agreeable fitting of experimental spectra. \cite{rg, pa, mcb}  This is the expected electronic configuration, arising from the hole doping introduced by the substitution of La$^{3+}$ with Sr$^{2+}$, which oxidizes a proportional fraction of Mn ions to maintain charge balance resulting in a mixed Mn$^{3+}/Mn^{4+}$ valence state. From the spectrum and peak positions, it is concluded that Mn is in the mixed +3 and +4 oxidation state.
The oxygen spectrum are shown in Fig. 3(d), where two peaks are observed at 532.3 eV and 533.5 eV where former is due to elemental peak and later is a surface defect contribution. These peak correspond to the -2 oxidation state of oxygen.\cite{ilyas3,ilyas4,ilyas5} From the detailed analysis of the XPS spectra, we concluded the following oxidation states La$^{3+}$, Sr$^{2+}$, Mn$^{4+}$, and O$^{2-}$. This suggests the crystallization of an ordered phase of epitaxially grown La$_{2/3}$Sr$_{1/3}$MnO$_3$ samples.

\subsection{Magnetic study}
 Fig. 4(a) shows isothermal field magnetization for LSMO/STO at 300 K, 150 K and 50 K, with the external magnetic field applied in-plane. The $M(H)$ loops exhibit well saturated behavior. The diamagnetic contribution from the STO substrate has been subtracted. Bulk LSMO has a Curie temperature ($T_c$) of approximately 370 K, whereas thin films of LSMO shows variation in $T_c$ depending on thickness.\cite{Urushibara,Millis} In this study, the thickness of LSMO was carefully chosen to ensure fully developed Ferromagnetic state. \cite{Huijben}. The $M(H)$ loop is symmetric around the x-axis, ruling out the presence of exchange bias. The remnant magnetization ($M_r$) and corecitivity ($H_c$) are higher at 50 K, as $M_r$ increases with decreasing temperature, which is a characteristic behavior of ferromagnets.\cite{Shalini}.

Bulk SrRuO$_3$ is ferromagnetic with $T_c$ $\sim$160 K, whereas epitaxial SrRuO$_3$ films undergo ferromagnetic phase transition at $\sim$ 150 K.\cite{Koster} The thickness of SrRuO$_3$ is chosen wisely so that it remains ferromagnetic below the temperature 150 K. \cite{ishigami} Inset of Fig 1(a) represents field dependent magnetization data for SrRuO$_3$film grown on SrTiO$_3$ substrate. The film exhibits a well-defined ferromagnetic response with clear magnetic hysteresis, confirming its intrinsic long-range ferromagnetic ordering. The hysteresis loop is symmetric with respect to the field axis, indicating the absence of exchange-bias effects. 

To further understand the magnetic behaviour at the interface of two different ferromagnets, field dependent magnetization data were collected for LSMO/SRO deposited on STO.  Fig 4(b) show $M(H)$ loops for LSMO/SRO on STO, at 300 K, 150 K and 50 K. As evident form Fig 4(b) the $M(H)$ loop at 300 K is dominated by LSMO. At 150 K, the loop exhibits lower coercivity ($H_c$) and remanence ($M_r$). However, below 150 K, $H_c$ increases drastically. This significant increase in ($H_c$) at 50 K, is quite interesting and may be attributed to interfacial effects. To further examine the magnetic behaviour at the interface of SRO and LSMO, the number of interfaces was increased from $n$ = 1 to $n$ = 5. Fig 4(c) shows isothermal field dependent magnetization data for (LSMO/SRO)$_5$ at temperatures 300 K, 200 K, 150 K, and 50 K. The measurements were performed with magnetic field parallel to film plane. At 300 K and 200 K, the hysteresis loop primarily reflects the magnetization of LSMO making it similar to Fig. 4(a). At 150 K, the hysteresis loop is remains closed. However at 50 K, the loop become more open, with higher values of $M_r$ = 72.07 emu/cm$^2$ and $H_c$ = 167 Oe as compared to 150 K. The hysteresis loop remains symmetric around $H$ = 0 showing no signature of exchange bias. This substantial opening of hysteresis loop at low temperature is particularly interesting.

The magnetic behavior of LSMO is known to be highly sensitive to epitaxial strain arising from lattice mismatch with the underlying substrate. This strain significantly influences the magnetic anisotropy of the film. Previous reports have demonstrated that compressive strain tends to favor out-of-plane magnetization by altering the orbital occupancy and enhancing the perpendicular anisotropy. In contrast, tensile strain promotes in-plane magnetization. Furthermore, it has been conclusively shown that under tensile strain conditions, the magnetic easy axis of LSMO lies within the plane of the film, reinforcing the strain-induced anisotropy effect.
In our present study, magnetization hysteresis measurements (as illustrated in Fig. 4(c)) reveal a distinct switching behavior of the magnetic layers in the $(LSMO/SrRuO_3)_5$ superlattice. When the external magnetic field is applied and swept in the positive direction, the magnetic moments in both LSMO and SRO layers align parallel to the field, indicating a ferromagnetic alignment under saturation. However, upon reducing the field from positive saturation, we observe that LSMO being magnetically softer than SRO, undergoes magnetization reversal at a lower coercive field. The reversal of the SRO layer follows at a higher field of $\sim$ 190 Oe. This sequential switching behavior strongly suggests the presence of interlayer magnetic coupling between the LSMO and SRO layers.
We propose that the observed reversal mechanism is driven by antiferromagnetic (AFM) coupling at the interface of LSMO and SRO layers. This AFM interaction is likely mediated through the Mn-O-Ru bond, where the magnetic moments of Mn and Ru ions couple antiparallel across the interface. Such a coupling mechanism introduces an additional energy barrier for the reversal of the SRO layer, causing it to switch at a higher field compared to LSMO. This explanation aligns well with previous literature, where similar interfacial AFM coupling in LSMO/SRO heterostructures has been reported. In one such study, temperature-dependent magnetization measurements further confirmed the existence of antiferromagnetic coupling at the interface, highlighting its role in modifying the overall magnetic response of the heterostructure. Despite the supporting evidence, a more comprehensive understanding of the interfacial magnetic interactions and their dependence on structural factors such as film orientation, thickness, and crystallinity is still required. \cite{Koster, Boschker, yichi}

\begin{figure}[h!]
	\centering
		\includegraphics[width=8cm]{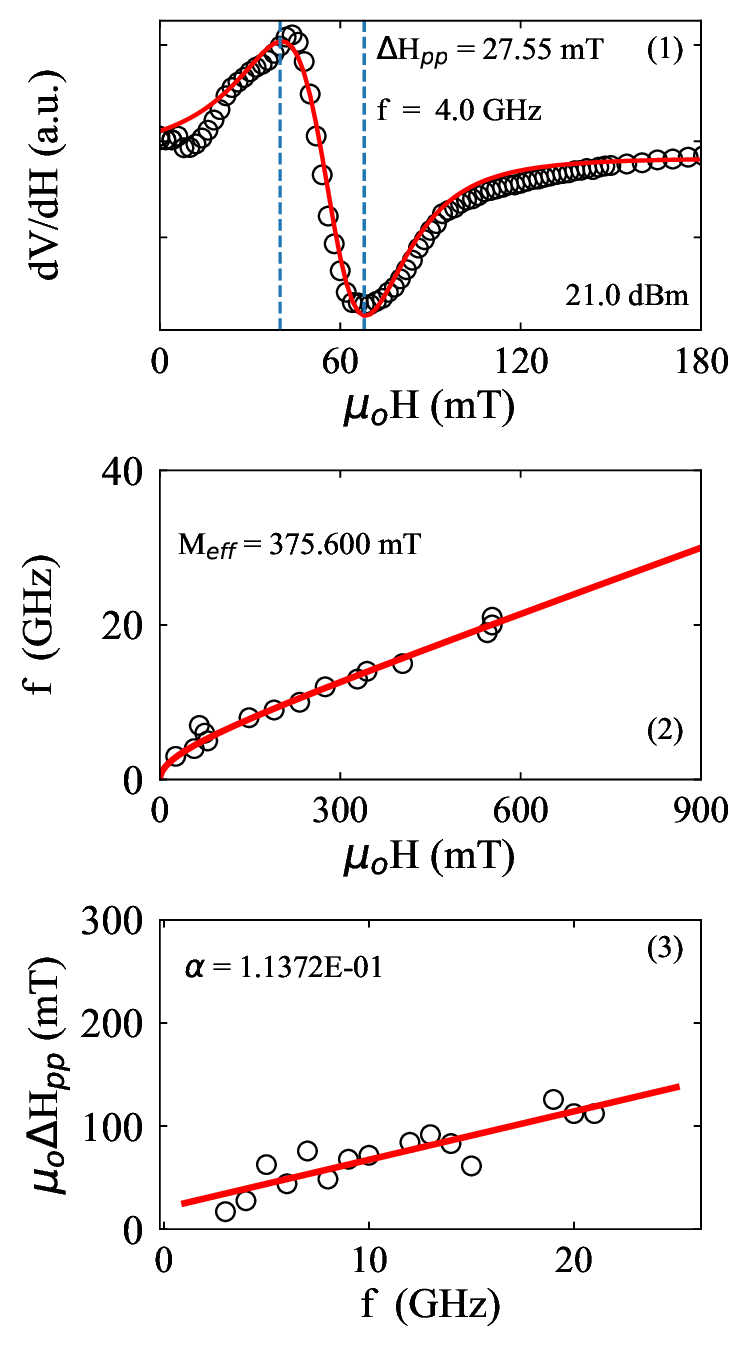}
	\caption{(color online) FMR analysis for single layer LSMO/STO(001) sample (1) FMR signal for 4.0 GHz at 21.0 dBm, solid red line is fitting due to sum of symmetric and anti-symmetric Lorentz function (2) resonance field plotted against resonance frequency, solid line is fitting due to Kittel's formula (see text) (3) shows linewidth ($\Delta$H$_pp$) as a function of resonance frequency, solid line is linear fitting (see text). The obtained values are inserted in the corresponding figures.}
	\label{fig:Fig5}
\end{figure}

\begin{figure}[h!]
	\centering
		\includegraphics[width=8cm]{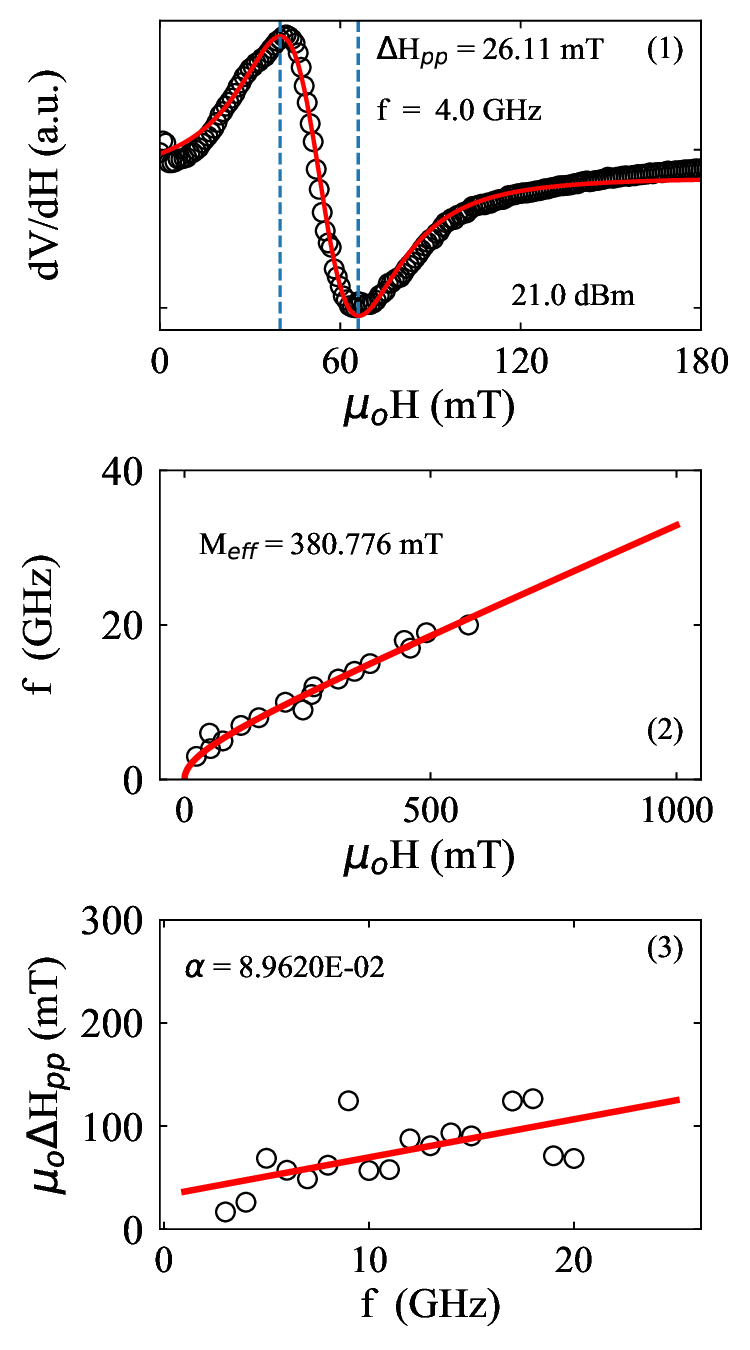}
	\caption{(color online) FMR analysis for superlattice (LSMO/STO)$_5$/STO(001) (1) FMR signal for 4.0 GHz at 21.0 dBm, solid red line is fitting due to sum of symmetric and anti-symmetric Lorentz function (2) resonance field plotted against resonance frequency, solid line is fitting due to Kittel's formula (see text) (3) shows linewidth ($\Delta$H$_pp$) as a function of resonance frequency, solid line is linear fitting (see text). The obtained values are inserted in the corresponding figures.}
	\label{fig:Fig6}
\end{figure}

\subsection{Microwave magnetic dynamics}
Magnetic dynamics of samples were investigated by broadband ferromagnetic resonance spectroscopy (FMR) presented for representative samples in Fig. 5 and 6. Fig. 5(1) and 6(1) shows a typical FMR spectra for LSMO and LSMO/SRO superlattice at 4.0 GHz. We have measure the FMR for all samples in a frequency range of 2-20 GHz at room temperature. The FMR absorption profile is well fitted with sum of symmetric and anti-symmetric Lorentz function\cite{gon} as shown in Figure 5(1) and 6(1) red solid line is due to fitting and we extracted resonance field H$_{res}$ for corresponding frequencies from fitting parameters, in fact we have also extracted the linewidth $\Delta$H of absorption peak.

The resonance frequency is plotted as a function on obtained resonance field as shown in Fig. 5(2) and Fig. 6(2) for both representative samples. The resonance frequency with in-plane external field is fitted with Kittel's formula given as $f = \frac{\gamma\mu_o}{2\pi}\sqrt{H_{ext}(H_{ext} + 4\pi M_{eff})}$, where $f$ is resonance frequency, $\gamma/4\pi$ = 0.028 GHz/mT and 4$\pi$M$_{eff}$ = 4$\pi$M$_s$ - H$_a$, H$_a$ =2K$_1$/M$_s$ is anisotropy field. The solid red line in Fig. 5(2) and Fig. 6(2) is fitting due to Kittel's formula, we obtained effective demagnetizing field (4$\pi$M$_{eff}$) as 375.6 mT 378.5 mT and 380.8 mT for LSMO, (LSMO/SRO)$_1$ and (LSMO/SRO)$_5$ respectively. The obtained values are in agreement with the earlier reported values.\cite{luo} Further the peak to peak linewidth ($\Delta H_{pp} = \Delta H /\sqrt{3}$) is plotted in Fig. 5(3) and 6(3). The FMR linewidth $\Delta$H is defined by a linear dependence on the resonance frequency $f$ and is connected with Gilbert damping constant ($\alpha$) by $\mu_o \Delta H = \mu_o \Delta H_o + (4 \pi \alpha_{eff)}/\gamma)f$. We have fitted the $\Delta H_{pp}$ as a function of frequency by the modified relation $\mu_o \Delta H_{pp} = \mu_o \Delta H_o + ((4 \pi \alpha_{eff})/(\gamma\sqrt{3})) f$ shown by solid line in Figs. 5(3) and 6(3) for LSMO and (LSMO/SRO)$_5$ superlattice respectively. The obtained values of gilbert damping constant are $1.11 \times 10^{-1}$, $9.87 \times 10^{-2}$ and $8.96 \times 10^{-2}$ for LSMO, (LSMO/SRO)$_1$ and (LSMO/SRO)$_5$ respectively.

\section{Summary and Conclusion}
Ferromagnetic half-metallic manganites are widely regarded as promising materials for spintronic device applications. In particular, the magnetic anisotropy of La$_{0.7}$Sr$_{0.3}$MnO$_3$ (LSMO) can be systematically tuned by engineering the Mn--O--Mn bond geometry through interfacial coupling with adjacent oxide layers. SrRuO$_3$ (SRO) is especially suitable for this purpose due to its excellent epitaxial compatibility with LSMO and its closely matched lattice parameters. Previous studies have demonstrated that the Mn--O--Ru bond at the interface mediates antiferromagnetic coupling between consecutive layers.\cite{solignac,ziese1,yichi} In multilayer heterostructures, interfacial effects are significantly enhanced as the number of interfaces increases. In our (LSMO/SRO)$_5$ multilayer system, we observe a distinct and unconventional two-step magnetization switching process. Upon reversing the magnetic field, the LSMO layers switch first, followed subsequently by the SRO layers, corresponding to the sequence:\[\text{LSMO}\uparrow\text{SRO}\uparrow \;\rightarrow\; \text{LSMO}\downarrow\text{SRO}\uparrow \;\rightarrow\; \text{LSMO}\downarrow\text{SRO}\downarrow.\] Notably, this two-step switching behavior is absent in the (LSMO/SRO)$_1$ bilayer. Similar unconventional switching phenomena at oxide interfaces have been reported previously, further supporting the interfacial origin of the observed behavior.\cite{solignac,halder}

We have grown epitaxial thin film of SrRuO$_3$ and La$_{2/3}$Sr$_{1/3}$MnO$_3$ on SrTiO$_3$ (001) substrate. The bi-layer and superlattice of LSMO/SRO have shown sharp interface and strained on STO substrate as revealed by x-ray diffraction and reciprocal space mapping on samples.Surface micrograph by AFM shows smooth surface with fractional nano-metric roughness which is also in good agreement with  x-ray reflectivity results. Interfacial magnetic coupling via Mn-Ru interaction at the interface predominantly observed by the feature of isothermal magnetization at low temperature for LSMO/SRO bi-layer and superlattice. We observe FMR signal for all samples at room temperature it is observed from FMR analysis that the damping decreased by an order of magnitude in heterostructure.  The interfacial magnetic interaction which often leads to anisotropic exchange interaction, exotic spin texture etc., and gives an additional degree of freedom to tune physical properties of functional materials. Interfaces and superlattice with such tunable properties are potential canidates for tunable properties for technological applications.

\section{Acknowledgment}
We acknowledge Laboratoire Albert Fert (UMR137) CNRS, Thales, Universit$\acute{e}$ Paris-Saclay for sample synthesis and measurement facilities.

\section{Conflict of interest}
Authors have no confilict of interest to declare.

\end{document}